\documentclass[%
preprint,
onecolumn,
superscriptaddress,
titlepage,
nofootinbib,
 amsmath,
 amssymb,
 aps,
]{revtex4-2}

\usepackage[english]{babel}
\usepackage[utf8]{inputenc}
\usepackage[T1]{fontenc}
\usepackage{commath}
\usepackage{slashed}

\usepackage{centernot}

\usepackage{graphicx}

\usepackage{booktabs}

\usepackage[pdftex, pdftitle={Article}, pdfauthor={Author}]{hyperref}

\usepackage{journals}

\usepackage{dcolumn}

\usepackage[capitalize]{cleveref}

\usepackage{bm}

\usepackage{xcolor}

\usepackage{soul}

\usepackage{multirow}

\newcolumntype{d}[1]{D{.}{.}{#1}}

\newcommand{\unit}[1]{%
    \ensuremath{\, \mathrm{#1}}}

\newcommand{\pder}[1]{\frac{\partial}{\partial #1}}
\newcommand{\ppder}[1]{\frac{\partial^2}{\partial #1^2}}

\newcommand{\mcH}{\mathcal{H}}
\newcommand{\mcK}{\mathcal{K}}
\newcommand{\mcL}{\mathcal{L}}
\newcommand{\mcM}{\mathcal{M}}
\newcommand{\mcP}{\mathcal{P}}
\newcommand{\mcQ}{\mathcal{Q}}
\newcommand{\mcW}{\mathcal{W}}

\newcommand{\rrsst}{r^2\sin^2\theta}

\begin{document}

\title{Oscillation dynamics of scalarized neutron stars}

\author{Christian J. Kr\"uger}
    \email{christian.krueger@tat.uni-tuebingen.de}
    \affiliation{Theoretical Astrophysics, IAAT, University of T\"ubingen, 72076 T\"ubingen, Germany}

\author{Daniela D. Doneva}
    \email{daniela.doneva@uni-tuebingen.de}
    \affiliation{Theoretical Astrophysics, IAAT, University of T\"ubingen, 72076 T\"ubingen, Germany}
    \affiliation{INRNE—Bulgarian Academy of Sciences, Sofia 1784, Bulgaria}

\date{\today}

\begin{abstract}
Scalar-tensor theories are well studied extensions of general relativity that offer deviations which are yet within observational boundaries. We present the time evolution equations governing the perturbations of a nonrotating scalarized neutron star, including a dynamic spacetime as well as scalar field within the framework of such scalar-tensor theories. We employ a theory that allows for a massive scalar field or a self-interaction term and we study the impact of those parameters on the non-axisymmetric $f$-mode. The time evolution approach allows for a comparatively simple implementation of the boundary conditions. We find that the $f$-mode frequency is no longer a simple function of the star's average density when a scalar field is present. We also evaluate the accuracy of different variants of the Cowling approximation commonly used in previous studies of neutron star oscillation modes in alternative theories of gravity and demonstrate that it can give us not only qualitatively correct results, but in some cases also good quantitative estimates of the oscillations frequencies.
\end{abstract}

\maketitle

\section{Introduction}
Gravitational wave asteroseismology is a powerful tool to study the internal structure of neutron stars through the observed gravitational wave signal emitted once various oscillation modes are excited by some astrophysical process \cite{Andersson:1997eq,Andersson:1997rn,1999LRR.....2....2K}. For this purpose a large number of studies were performed examining the different classes of oscillation modes both for static \cite{Andersson:1997rn,Benhar:2004xg,Lau:2009bu,Blazquez-Salcedo:2013jka,Chirenti:2015dda} and rotating stars \cite{Font:2000rd,Gaertig:2008uz,Kruger:2009nw,Doneva:2013zqa,Kruger:2019zuz,Kruger:2020ykw}, including spacetime modes \cite{Kokkotas:2003mh,Andersson:1996ua}. While the effect of different equations of state, composition, temperature profile, differential rotation, etc. was studied in detail, little has been done in quantifying another source of uncertainty, that is the underlying theory of gravity. 

Within the framework of alternative theories of gravity, perhaps the most widely studied neutron star models were those in a class of scalar-tensor theories (STT) of gravity admitting the so-called scalarization. The reason is first that STT are some of the most natural and unproblematic extensions of general relativity (GR). Second, scalarization is a nonlinear effect allowing to endow highly compact objects with scalar hair while leaving the weak-field regime equivalent to GR and thus fully in agreement with the observations. Scalarization of neutron stars was first considered in the original work of Damour and Esposito-Farese \cite{Damour:1993hw}, extended later to slow \cite{Damour:1996ke,Sotani:2012eb,Pani:2014jra} and rapid rotation \cite{Doneva:2013qva}. Scalarized stars with anisotropic pressure were studied in \cite{Silva:2014fca} while the effect of magnetic field was examined in \cite{Soldateschi:2020hju,Soldateschi:2020zxb}. The most stringent constraint to date on this class of theories comes from the observations of pulsars in close binary systems and they limit the possible deviations from GR to a very small value \cite{Damour:1996ke,Freire:2012mg,Antoniadis:2013pzd,Shao:2017gwu}. One possibility to evade these constraints is to consider a nonzero scalar field mass that introduces a characteristic radius of the scalar field, associated with its Compton wavelength, beyond which the scalar field drops to zero exponentially \cite{Popchev2015,Ramazanoglu:2016kul,Doneva:2016xmf,Rosca-Mead:2020bzt}. In addition, a self-interaction term in the scalar field potential can have a qualitatively similar effect \cite{Staykov:2018hhc}. Neutron star mergers in such theories were considered in \cite{Sagunski:2017nzb} while the core-collapse was examined in \cite{Sperhake:2017itk,Cheong:2018gzn,Rosca-Mead:2019seq,Geng:2020slq}. An alternative way to evade the binary pulsar constraints is to consider for example multiple scalar fields \cite{Doneva:2020afj}.

The first study of neutron star oscillations in alternative theories of gravity was performed for scalarized neutrons stars in STT---the polar fluid modes were examined in \cite{Sotani:2004rq} while the spacetime $w$-modes were considered in \cite{Sotani:2005qx}. Calculating the polar perturbations is, in general, a much more involved task which is why the (which we will later dub ``full'') Cowling approximation was employed in \cite{Sotani:2004rq}, i.e. assuming that the spacetime as well as the scalar field are fixed and only the fluid perturbations are evolved. Even though this seems like a crude approximation, similar approaches have proven to give qualitatively good results in general relativity \cite{1990ApJ...348..198L,Gaertig:2008uz,Sotani:2020mwc} and that is why it is reasonable to adopt it as a first approximation to study the leading order effects of STT. The results in \cite{Sotani:2004rq} were later generalized to the case of rapid rotation \cite{Yazadjiev:2017vpg} where also the effect of the scalar field on the Chandrasekhar-Friedman-Schutz \cite{Chandrasekhar:1992pr,Friedman:1978hf} instability was examined in detail. The polar modes in the Cowling approximation in $f(R)$ gravity, which is mathematically equivalent to a particular class of STT \cite{Sotiriou:2008rp}, were considered in \cite{Staykov:2015cfa}.

The field developed further in the direction of calculating the axial modes in alternative theories of gravity \cite{Blazquez-Salcedo:2015ets,Blazquez-Salcedo:2018tyn,Blazquez-Salcedo:2018qyy,AltahaMotahar:2018djk}. Torsional oscillations of scalarized neutron stars were considered in \cite{Silva:2014ora}.  Another major effort was the calculation of radial modes which was first approached by keeping the spacetime metric fixed but allowing for the evolution of the fluid and the scalar field \cite{Sotani:2014tua}. The full problem without approximation was addressed in \cite{Mendes:2018qwo} where not only the stability of neutron stars in STT against small perturbations was proved but also the emergence of a new class of modes associated with the scalar field was demonstrated. These are the breathing modes that can be excited in processes such as core-collapse \cite{Gerosa:2016fri}. The first study of non-radial neutron star polar modes with $\ell \ge 2$ without approximation in alternative theories of gravity was performed in \cite{Blazquez-Salcedo:2020ibb}. There, the fundamental $f$-mode (which is the lowest frequency mode in barotropic perfect fluids having no nodes in radial direction) and its overtones (also called $p$-modes having one or several nodes) were calculated for compact objects in a special class of scalar-tensor theory with a massive scalar field which is mathematically equivalent to $R^2$ gravity.

In the present paper, we will concentrate on studying the polar oscillations modes, and more specifically the $\ell=2$ $f$-mode, of scalarized neutron stars in STT. We focus on the cases both with and without scalar field potential, even though only the former one can give us large deviations from GR if one considers values of the parameters in agreement with the observations. The oscillation modes are calculated by evolving the relevant perturbation equations in time. Even though this method is inferior in accuracy compared to solving the eigenvalue problem, the treatment of the boundary conditions is considerably simpler and allows for straightforward calculations both in the case of massive and massless scalar field. 

The paper is organized as follows. In Sec.~\ref{sec:formulation} the formulation of the problem and the basic equations are discussed. The perturbations equations governing the neutron star oscillations in STT are given in a separate appendix. The results are presented in Sec.~\ref{sec:results}. The paper ends with Conclusions.

Unless otherwise noted, we work in units in which $c=G=1$.

\section{Mathematical Formulation}
\label{sec:formulation}

\subsection{Background neutron star solutions in scalar-tensor theories}
The action of scalar-tensor theories is given in the Einstein frame by
\begin{align}
\label{eq:action}
    S
        & = \frac{1}{16\pi}
            \int \dif^4 x \sqrt{-g}
            \left[
                R - 2 g^{\mu\nu} \nabla_\mu \varphi \nabla_\nu \varphi
                - 4 V(\varphi)
            \right]
            + S_{\rm matter}\left(A^2(\varphi) g_{\mu\nu}, \chi\right),
\end{align}
where $R$ and $\nabla_\mu$ are the Ricci scalar and the covariant derivative with respect to the Einstein frame metric $g_{\mu\nu}$. $ V(\varphi)$ is the scalar field potential and $S_{\rm matter}$ is the action of the matter sources collectively denoted by $\chi$. The Einstein frame and the Jordan (physical) frame are linked via the conformal factor $A(\varphi)$, with the Jordan frame metric $\tilde{g}_{\mu\nu} = A^2(\varphi) g_{\mu\nu}$. Due to the simplicity of the field equations in the Einstein frame we will adopt it in our calculations and refer to the physical Jordan frame mainly for the final observable quantities that we calculate. Unless otherwise specified, the quantities in the Jordan frame will be marked with a tilde. A detailed description of the two frames and the transformations between both especially in the case of scalarized neutron stars can be found for example in \cite{Doneva:2013qva}

The field equations resulting from the action \eqref{eq:action} are
\begin{align}
    \label{eq:fieldeq}
    R_{\mu\nu} - \frac{1}{2} g_{\mu\nu} R
        & = 8\pi T_{\mu\nu} + 2 \nabla_\mu \varphi \nabla_\nu \varphi
            - g_{\mu\nu} g^{\alpha\beta} \nabla_\alpha\varphi \nabla_\beta\varphi
            - 2V(\varphi) g_{\mu\nu},
    \\
    \label{eq:fieldeqscalar}
    \nabla^\mu \nabla_\mu \varphi
        & = 
            - 4\pi \alpha(\varphi) T + \frac{\partial V(\varphi)}{\partial\varphi},
\end{align}
where we have defined the coupling function $\alpha(\varphi) := \frac{\dif\, \ln A(\varphi)}{\dif\varphi}$. By using the contracted Bianchi identities, we find that the conservation law for the energy-momentum tensor in the Einstein frame takes the form
\begin{align}
    \label{eq:convlaw}
    \nabla_\mu T^\mu {}_\nu 
        & = \alpha(\varphi) T \nabla_\nu \varphi,
\end{align}
where $T$ is the trace of the energy-momentum tensor.

We will work with a perfect fluid neutron star, for which the energy-momentum tensor (in the Einstein frame) is, as usual, given by
\begin{align}
    T_{\mu\nu}
        & = (\epsilon + p) u_\mu u_\nu + p g_{\mu\nu},
\end{align}
where $\epsilon$ is the energy density, $p$ the pressure, and $u_\mu$ the four-velocity of the fluid.

In order to calculate the background neutron star models and their oscillation spectrum, we have to employ an equation of state $\tilde{p} = \tilde{p}(\tilde{\epsilon})$ which will be provided in the physical Jordan frame. For this purpose we need relations between the fluid quantities in the two frames. Given that $T_{\mu\nu} = A^2(\varphi) \tilde{T}_{\mu\nu}$, one can easily show that for a perfect fluid $\epsilon = A^4(\varphi) \tilde{\epsilon}$, $p = A^4(\varphi) \tilde{p}$, and $u_\mu = A^{-1}(\varphi) \tilde{u}_\mu$. Since the energy density and pressure are the only two fluid quantities which we need to know in the Jordan frame, we will transform only those two to the Jordan frame when necessary but will otherwise work with the Einstein frame.

In the present study, we will consider non-rotating neutron stars, for which the line element in isotropic coordinates can be written as
\begin{align}
    \dif s^2
    & = - e^{2\nu} \dif t^2
        + e^{2\psi} \dif r^2
        + e^{2\psi} r^2 \dif \Omega^2
\end{align}
where $\nu$ and $\psi$ are the two metric potentials and $\Omega$ is the solid angle. As a result, the four-velocity of the fluid is given by $u^\mu = ( e^{-\nu}, 0, 0, 0)$. The dimensionally reduced field equations assuming the above form of the metric, can be found in \cite{Yazadjiev:2016pcb}.

We restrict ourselves to the study of the dynamics of small perturbations around an equilibrium configuration for which we have to linearise Eqs. \eqref{eq:fieldeq} to \eqref{eq:convlaw}. We will use the same formalism that has been presented in Ref.~\cite{Kruger:2020ykw}; nonetheless, not only for completeness, but also since we use merely the ``nonrotational subset'' of the formalism presented there (however, extended for the presence of a scalar field $\varphi$), we will repeat the basics for clarity.

We will introduce time-dependent perturbations for which we will derive evolution equations. First, we decompose the metric as
\begin{align}
    g_{\mu\nu}
        & = g_{\mu\nu}^{(0)} + h_{\mu\nu},
\end{align}
where $g_{\mu\nu}^{(0)}$ is the background metric and $h_{\mu\nu}$ its perturbation; we use the background metric to raise and lower the indices of the latter. As we will employ the Hilbert gauge, it will be advantageous to work instead with the trace-reversed metric perturbation, defined by
\begin{align}
    \phi_{\mu\nu}
        & := h_{\mu\nu} - \frac{1}{2} g_{\mu\nu}^{(0)} h,
\end{align}
where $h := {h^\mu}_\mu$ is the trace of the metric perturbations. The Hilbert gauge, which is the gravitational equivalent to the well-known Lorenz gauge in electromagnetism, is specified by
\begin{equation}
    f_\mu := \nabla^\nu \phi_{\mu\nu} = 0.
    \label{eq:hilbert_gauge}
\end{equation}
We opt for this gauge as it provides us directly with wave equations for the spacetime perturbations, which can, without further ado, be used for a time evolution (see Ref.~\cite{Kruger:2020ykw} for a more in-depth discussion). As we consider perturbations on a spherically symmetric background, we will separate out the angular dependence of the perturbation variables by means of the spherical harmonics (and their vector and tensor counterparts).

The precise definition of the perturbation variables for the spacetime, the fluid and the scalar field can be found in Appendix~\ref{app:def_pert} while the resulting set of evolution equations is stated Appendix~\ref{app:evol_eq}.

\subsection{Fixing the theory}
Let us discuss the particular class of STT we will concentrate on. The free functions that define the theory are the conformal factor $A(\varphi)$ (or equivalently the function $\alpha(\varphi)$) and the scalar field potential $V(\varphi)$. The simplest assumption we can make for the coupling $\alpha(\varphi)$ is that it can be expanded in series with respect to $\varphi$. Keeping only terms up to linear order in $\varphi$ and assuming that the cosmological background value of the scalar field $\varphi_0=0$, we have
\begin{equation}
    \alpha(\varphi)=\alpha_0 + \beta \varphi.
\end{equation}
The case with $\alpha_0\ne 0$ and $\beta=0$ is equivalent to the famous Brans-Dicke theory that is, though, severely constrained by the weak field observations \cite{Will:2014kxa}. In the present paper we will focus on the Damour-Esposito-Far\'ese (DEF) model \cite{Damour:1993hw} with $\alpha_0 = 0$ and $\beta \ne 0$ that is perturbatively equivalent to GR in the weak field regime but can lead to nonlinear development of the scalar field for highly compact objects such as neutron stars, that is the so-called scalarization. Such scalarization can happen for negative $\beta$ and the more negative $\beta$ is, the stronger is the scalar field (for scalarization in special cases with large positive $\beta$ we refer the reader to \cite{Mendes:2014ufa,Mendes:2016fby,Mendes:2019zpw}). Binary pulsar observations set very tight constrains on the lower limit of $\beta$, more precisely $\beta>-4.5$ \cite{Damour:1996ke,Freire:2012mg,Antoniadis:2013pzd,Shao:2017gwu}. This criterion is dependent on the equation of state, but taking into account that scalarization may happen for roughly $\beta<-4.35$ \cite{Damour:1996ke}, it is evident that there is little room for deviations from GR. 

As discussed in the introduction, the scalar field mass term (and possibly a self-interaction one) in the potential can help us evade these constraints by introducing a finite range of the scalar field of the order of its Compton wavelength. In accordance with \cite{Doneva:2016xmf,Staykov:2018hhc}, we will consider the following form of $V(\varphi)$
\begin{equation} \label{eq:potential}
    V(\varphi) = \frac{1}{2} m_\varphi^2 \varphi^2 + \frac{1}{4}\lambda \varphi^4,
\end{equation}
where $m_\varphi$ is the scalar field mass and $\lambda$ the self-interaction parameter. As already stated, we use units in which $c=G=1$. Hence, in the way the potential is defined, $m_\varphi$ has dimensions of $L^{-1}$ (where $L$ is a length) while $\lambda$ has dimensions of $L^{-2}$. That is why we have introduced a dimensionless scalar field mass and self-interaction parameter defined as $m_\varphi \rightarrow m_\varphi R_0$ and $\lambda \rightarrow \lambda R_0^2$, where $R_0 = 1.47664\unit{km}$ is one half of the solar gravitational radius.

\subsection{Numerical Implementation}

The numerical implementation of the evolution equations follows what is described in the previous paper~\cite{Kruger:2020ykw} but is considerably less involved since we are dealing here with a purely radial problem.\footnote{Our problem is \emph{radial} in the sense that our time evolution equations have no explicit angular dependence as it is completely separated out by means of the spherical harmonics. This should not be confused with \emph{radial modes} which have actually no angular dependence themselves; our study is indeed concerned with nonaxisymmetric modes.} We will briefly recall the relevant details; for further information, we refer the reader to~\cite{Kruger:2020ykw}.

The numerical treatment of the evolution equations remains essentially the same: we use finite differences to discretize spatial derivatives on the same radial grid and employ a 3rd order Runge-Kutta scheme for time integration; we use Kreiss-Oliger dissipation to damp out numerical instabilities \cite{kreiss1973methods}. The boundary conditions of the fluid perturbations at the surface of the star and those of the space-time perturbations at the outer boundary of the numerical grid are identical to those in \cite{Kruger:2020ykw}. We treat the scalar field perturbation at the outer boundary identical to the space-time perturbation (by applying an outgoing-wave boundary condition).

The boundary conditions at the origin deserve an extra comment since we have separated out the spherical harmonics from our perturbation variables (see App.~\ref{app:def_pert} for their definitions) and, subsequently, our evolution equations (see App.~\ref{app:evol_eq}) now involve the order number $\ell$ rather than the azimuthal parameter $m$. After expanding the evolution equations around the origin $r=0$ in a Taylor expansion, we find that the perturbation variables $\{\mcH_0, \mcL_0, \mcM_0, Q_{10}, Q_{30}, Q_{40}, \delta\varphi_0\}$ have to be zero at the origin, while the other four variables $\{\mcK_0, \mcP_0, \mcQ_0, \mcW_0\}$ need to have a vanishing radial derivative there.

\subsection{Initial Data}

The solution of a time evolution problem requires the specification of initial data which are then evolved in time; the result obviously depends fundamentally on this choice. Ideally, one would specify initial data that represent a certain astrophysical scenario and the time evolution then reveals the future development of the system.

However, devising such initial data is a very complex task by itself; furthermore, we are not interested in the particular time evolution of the system but rather in its spectrum of vibrations which are characterized solely by a frequency and a damping time. Any given set of initial data can be decomposed into a superposition of the system's eigenfunctions whose evolution then is---in a linear system like ours---independent from each other. In this way, depending on which particular initial data we choose, we will excite different eigenmodes of the system with different amplitudes. A Fourier analysis will then unveil the frequencies and amplitudes of the individual modes that have been excited.

As our main interest lies in the $f$-mode of the neutron star fluid, we will devise initial data that roughly resemble its eigenfunction in order to excite it strongest. We will not be able (ultimately due to numerical truncation error), and it is also not our aim, to excite merely the $f$-mode and no other modes at all. In general, other modes like its overtones will also be excited but with a smaller amplitude.

In order to excite mainly the $f$-mode, we will prescribe a nonzero function for $Q_3$, which is closely related to the radial velocity perturbation; we set all other perturbation variables to zero on the initial time slice. In particular, we set
\begin{align}
    Q_3(t=0, r)
        & = \sqrt{r} \left( r_e - r \right)\quad\text{for}\quad r < r_e,
\end{align}
where $r$ is the coordinate radius and $r_e$ its value at the surface of the star. This function very roughly (up to about 30\% error) approximates the $f$-mode eigenfunction, i.e., we will inevitably excite other fluid modes as well; as explained above, this does not pose any problem for our study.

\section{Results}
\label{sec:results}

The quantities that will be presented in this section will of course be in the physical Jordan frame. The mass definition in scalar-tensor theories of gravity is a subtle problem and it turns out that only the Einstein frame ADM mass has natural energy-like properties (see, e.g., the discussion in \cite{Doneva:2013qva} and references therein), and that is why we will use it for the plots. The physical circumferential radius can be obtained as
\begin{equation}
    \tilde{R}_c = R_{\rm EF} A(\varphi),
\end{equation}
where $R_{\rm EF}$ is the Einstein frame radius at which the pressure vanishes. The oscillation frequencies, on the other hand, are the same in both frames \cite{Yazadjiev:2017vpg}.

\subsection{Background models}

\begin{figure}
	\includegraphics[width=0.70\textwidth]{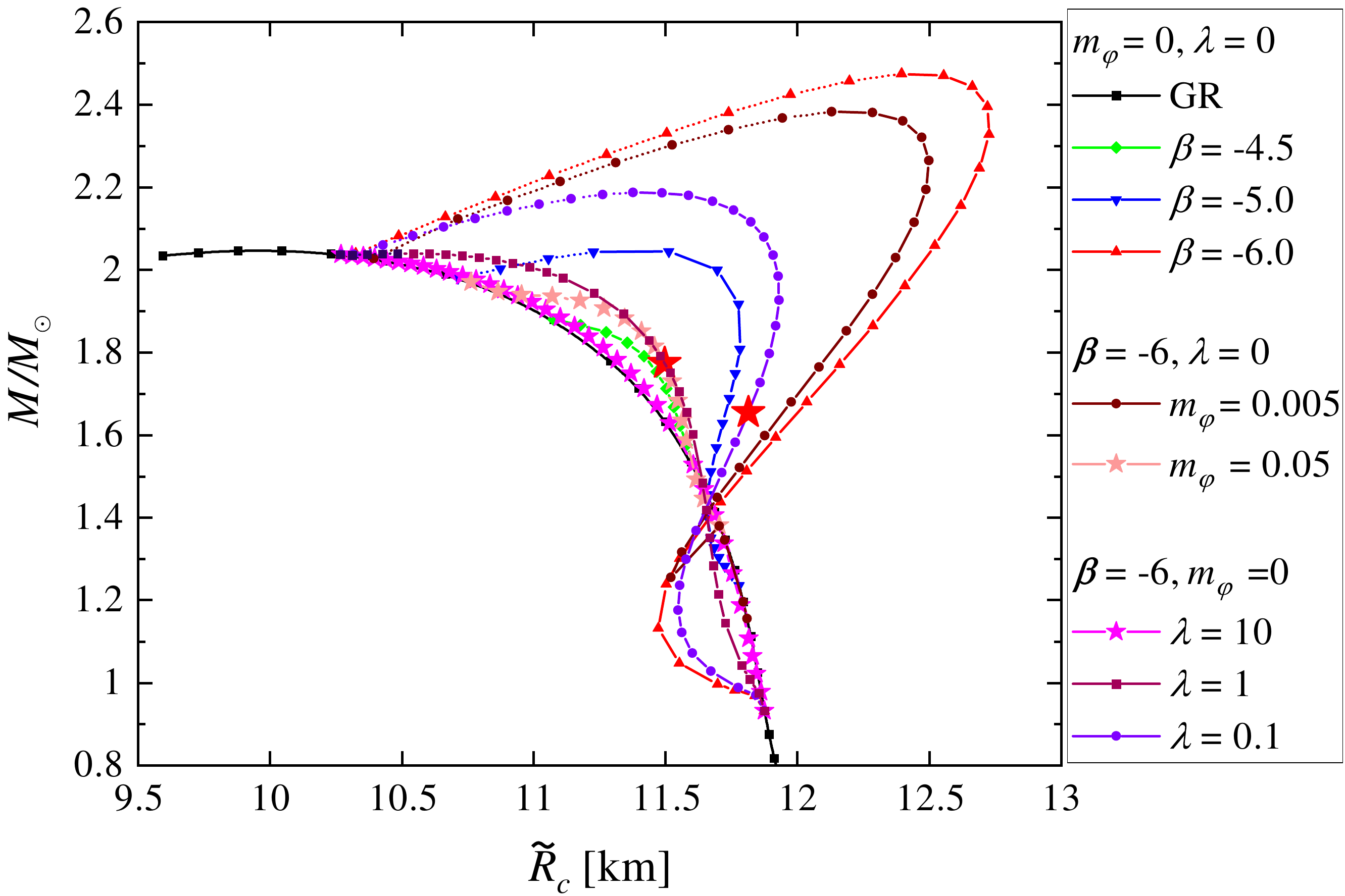}
	\caption{The mass as a function of the radius for all of the sequences we employed. The two models for which we show the time evolution in Figs.~\ref{fig:timepsd1} and \ref{fig:timepsd2} are marked with large red stars in the figure. }
	\label{fig:M_R}
\end{figure}

Let us first discuss the sequences of background neutron star models that we have constructed in order to study their dynamics. Our main aim is to explore in detail the behavior of the oscillation frequency with the change of the different parameters of the theory. This is complicated by the fact that we have three independent parameters coming only from the STT, that is the coupling parameter $\beta$, the scalar field field mass $m_\varphi$, and the self-interaction parameter $\lambda$. There is one more free input in the problem that is the equation of state (EOS). We have decided to limit our studies to only one modern realistic EOS in order to make the presentation of the results more tractable, the SLy EOS \cite{2001A&A...380..151D}; in fact, we use its piecewise-polytropic approximation \cite{2009PhRvD..79l4032R} for the ease of numerical implementation. As far as the STT parameters are  concerned we have considered a large range of parameters and the sequences used for the time evolution are shown in a common mass-radius diagram in Fig.~\ref{fig:M_R} (the black curve with filled squares represents the well-known mass-radius curve in pure GR). For each of the sequences the part that is before the maximum of the mass is plotted with a solid line, while the unstable one is plotted with a dotted line. The same convention is used also in most of the figures below.

The coupling parameter $\beta$ spans the interval from $-4.5$, which is marginally in agreement with binary pulsar observations, until $-6$. While $\beta<-4.5$ is not allowed by the observations in the massless case, this changes dramatically for even very small scalar field mass or small values of $\lambda$. The qualitative effect of nonzero $m_\varphi$ and $\lambda$ is quite similar -- the scalar field is suppressed leading to smaller deviations from GR. While $\lambda \ne 0$ preserves the position of the bifurcation points (for a fixed $\beta$), the range of central energy densities where scalarization occurs shrinks with increasing scalar field mass $m_\varphi$. As it was demonstrated in \cite{Danchev:2020zwn,Ramazanoglu:2016kul,Yazadjiev:2016pcb}, $m_\varphi \sim 10^{-13}{\rm eV}$ (or equivalently $m_\varphi \sim 10^{-2}$ in our dimensionless units) will lead to a complete suppression of the scalar dipole radiation in the binary pulsar observations, while the corresponding neutron star models are practically indistinguishable in their bulk properties, such as mass and radius, from the corresponding model with the same $\beta$ but $m_\varphi=0$. That is why the case of $m_\varphi = 0$ and $\lambda = 0$ results in the maximum possible deviation from GR for a fixed value of $\beta$ (even though it is rigorously speaking in contradiction to observations). In the case of nonzero scalar field potential, the presented sequences are for $m_\varphi\ne 0$ and $\lambda=0$, and for $m_\varphi = 0$ and $\lambda \ne 0$ in order to better distinguish between effects coming from the two different terms in Eq.~\eqref{eq:potential}.

\subsection{Oscillation spectrum}

We perform a time evolution of the perturbation quantities and place an observer at some location inside the star. Our simulations usually cover around $8\unit{ms}$ and the interior of the star is well resolved by 240 grid points. As we are interested in the fluid modes, we use as initial data a perturbation that roughly resembles that of an $f$-mode eigenfunction of one of the fluid quantities. The power spectral density (henceforth PSD) of the time series taken by the observer then reveals the $f$-mode as well as a few of its overtones (the pressure modes $p_n$).

\begin{figure}[htpb]
	\includegraphics[width=1.00\textwidth]{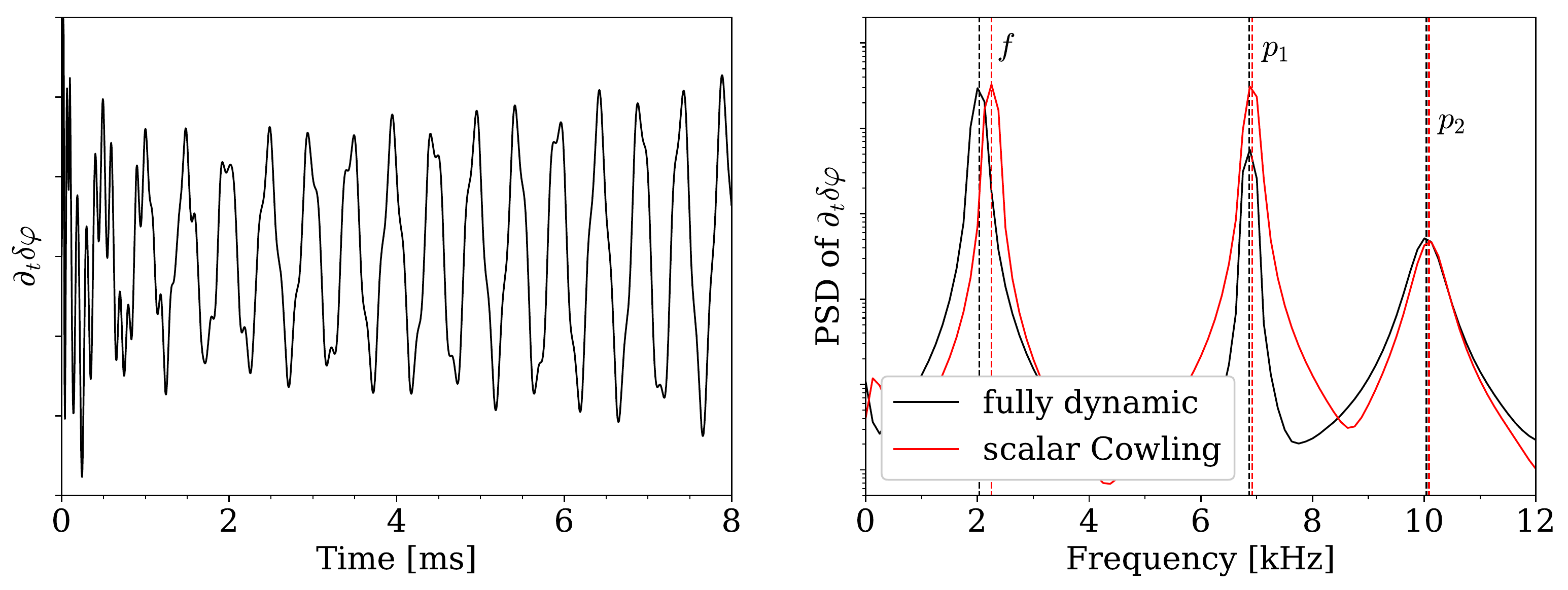}
	\caption{Time series (left panel) and PSD (right panel) of the model with $\epsilon_c = 1.3 \times 10^{15} \text{g/cm}^3$, $\beta = -6$, $m_\varphi = 0$, and $\lambda = 0.1$. Shown is the time derivative of the scalar field perturbation. The black PSD corresponds to the fully dynamic case (shown in the left panel), whereas the red PSD is obtained while keeping the scalar field fixed; this approximation slightly increases the fluid mode frequencies. The $f$-mode (at $2.034\unit{kHz}$ and $2.249\unit{kHz}$, respectively) and its first two overtones $p_1$ and $p_2$ are clearly visible.}
	\label{fig:timepsd1}
\end{figure}

As a first characteristic example, we discuss the simulation of a scalarized neutron star model with massless scalar field and non-zero self-interaction parameter $\lambda$; we show the time series of the time derivative of the scalar field perturbation, $\partial_t \delta\varphi$, in the left panel of Fig.~\ref{fig:timepsd1}. The PSD of that time series is shown in the right panel (in black) and the $f$-mode and its first two overtones are clearly distinguishable (their frequencies are independent of which perturbation variable we analyse); as we started with initial data resembling an $f$-mode eigenfunction, higher overtones at even higher frequencies are only weakly excited and not as clearly visible. In the same panel, we show in red the PSD of the time series of the same stellar model, however, this time while keeping the scalar field fixed (``scalar Cowling''), i.e., $\delta\varphi = 0$, that will be discussed in detail in the Sec.~\ref{sec:cowling}. The numerical values of the frequencies are shown in the top half of Tab.~\ref{tab:freq}.

\begin{table}[htpb]
    \centering
    \caption{The frequencies of the $f$-mode and its first two overtones of the two models shown in Figs.~\ref{fig:timepsd1} (top half) and \ref{fig:timepsd2} (bottom half); see those captions for details on the models. The first row contains the frequencies in the fully dynamic case, whereas the second row contains the slightly increased frequencies due to keeping a degree of freedom fixed (see text for further detail). All frequencies are given in kHz. The naming convention of the different Cowling approximations will be introduced in Sec.~\ref{sec:cowling}.}
    \label{tab:freq}
    \begin{tabular}{ccccc}
        \toprule
        STT parameters
            & evolution type & $f$ & $p_1$ & $p_2$ \\
        \midrule
        \multirow{2}{*}{$m_\varphi = 0$, $\lambda = 0.1$ (cf. Fig.~\ref{fig:timepsd1})}
            & fully dynamic & 2.034 & 6.867 & 10.040 \\
            & ``scalar Cowling'' & 2.249 & 6.918 & 10.085 \\
        \midrule
        \multirow{2}{*}{$m_\varphi = 0.05$, $\lambda = 0$ (cf. Fig.~\ref{fig:timepsd2})}
            & fully dynamic & 2.122 & 6.571 & 9.644 \\
            & ``classic Cowling'' & 2.552 & 7.300 & 10.050 \\
        \bottomrule
    \end{tabular}
\end{table}

\begin{figure}
	\includegraphics[width=1.0\textwidth]{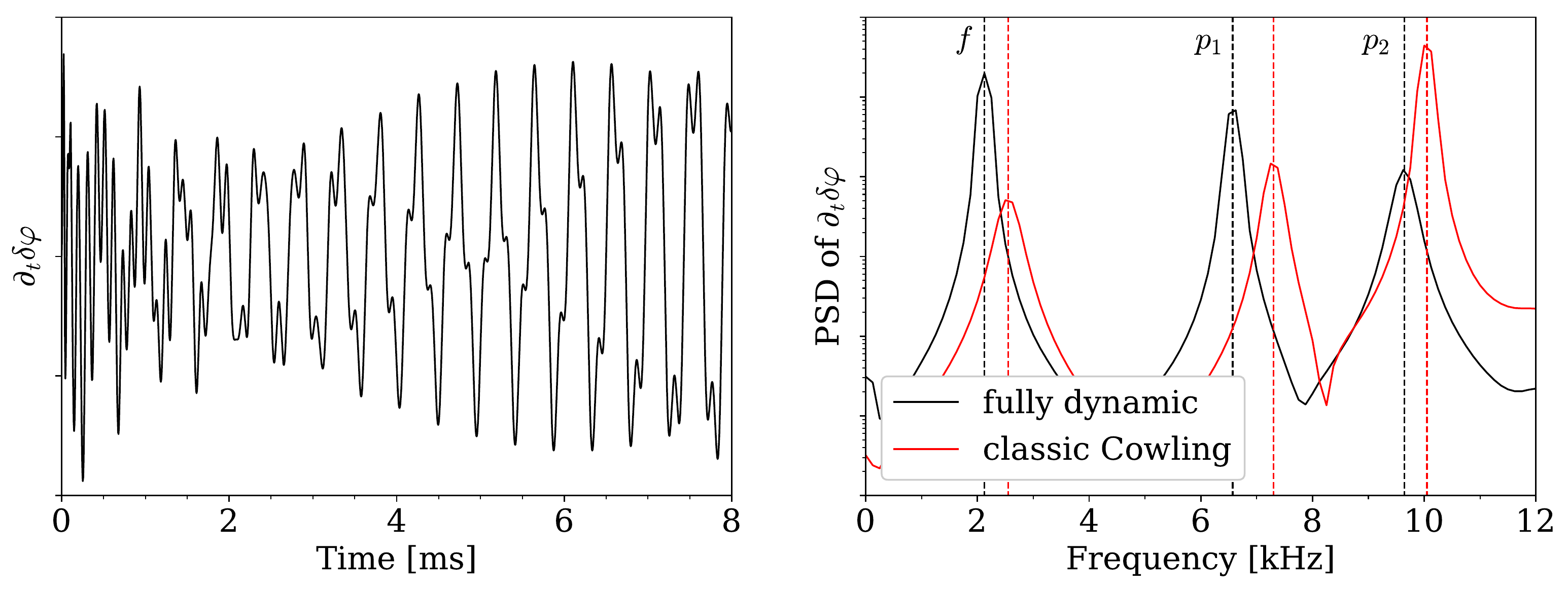}
	\caption{Time series (left panel) and PSD (right panel) of the model with $\epsilon_c = 1.4 \times 10^{15} \text{g/cm}^3$, $\beta = -6$, $m_\varphi = 0.05$, and $\lambda = 0$. Shown is the time derivative of the scalar field perturbation. The black PSD corresponds to the fully dynamic case (shown in the left panel), whereas the red PSD is obtained while keeping the space-time fixed; the Cowling approximation increases the fluid mode frequencies. The $f$-mode (at $2.122\unit{kHz}$ and $2.552\unit{kHz}$, respectively) and its first two overtones $p_1$ and $p_2$ are clearly visible. The wider Fourier peaks and the apparently increased energy stored in the overtones is an artifact of the Cowling approximation. }
	\label{fig:timepsd2}
\end{figure}

In Fig.~\ref{fig:timepsd2}, we display another exemplary time series and its PSD (in black), this time for a scalarized neutron star model in which the scalar field has non-zero mass but instead the self-interaction parameter $\lambda$ of the potential is zero. Again, we show in red for comparison the PSD of the same model, however, this time with the space-time perturbations held constant and allowing for evolution of $\delta \varphi$ (``classic Cowling''), that will be discussed in detail in the following Sec.~\ref{sec:cowling}. The precise frequencies of the visible peaks can be found in the bottom half of Tab.~\ref{tab:freq}. 

Our extracted frequencies are accurate to 1-2\%; we have established this accuracy of our code in our previous study \cite{Kruger:2020ykw} and also by running a few of the present simulations at different resolutions. Furthermore, our results are in very good agreement with the results by Bl\'azquez-Salcedo et al.~\cite{Blazquez-Salcedo:2020ibb}.

\begin{figure}
	\includegraphics[width=1.1\textwidth]{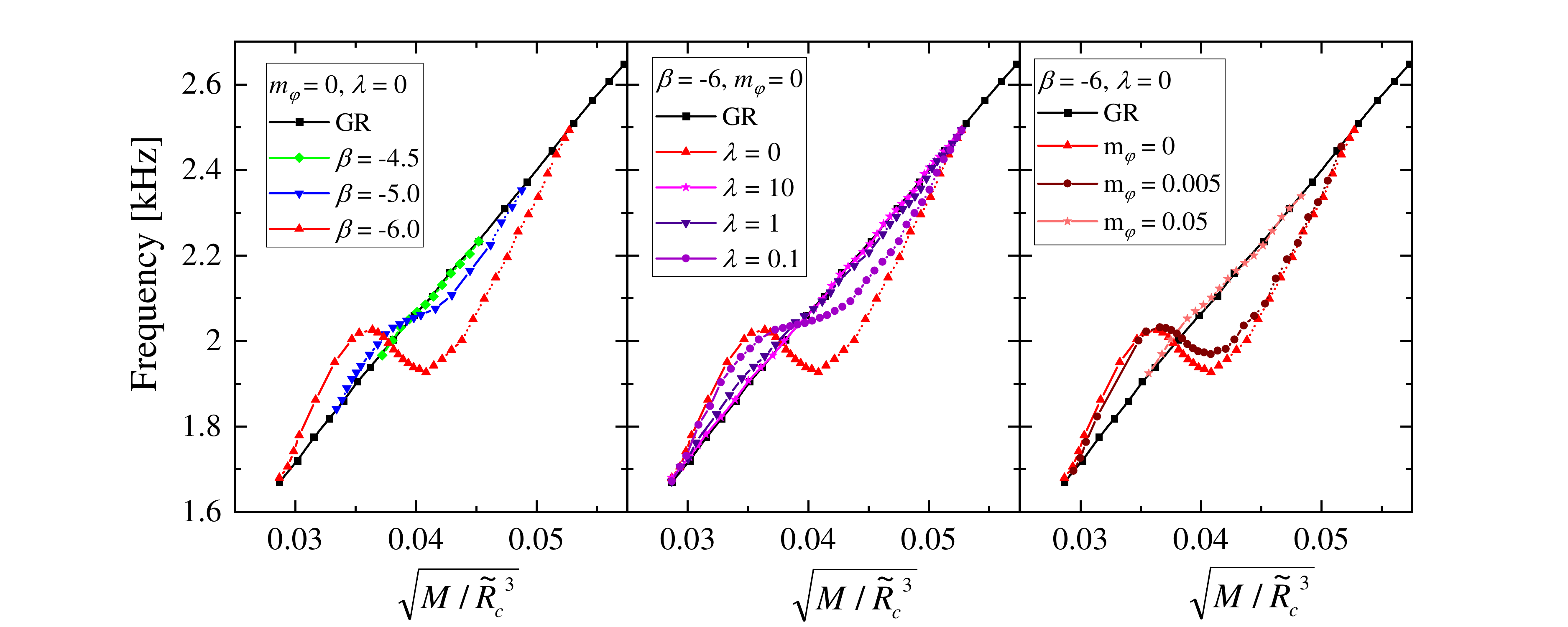}
	\caption{The $f$-mode frequency as a function of the average density $\sqrt{M/\tilde{R}_c^3}$. \textit{(Left panel)} $m_\varphi = 0$, $\lambda = 0$ and $\beta$ is varied.\textit{(Middle panel)} $\beta=-6$, $m_\varphi = 0$ and $\lambda$ is varied.  \textit{(Right panel)} $\beta = -6$, $\lambda = 0$ and $m_\varphi$ is varied.  }
	\label{fig:f_MR3}
\end{figure}

Next, we will turn to the $f$-mode frequency and study how it is impacted by the different scalar field parameters. We start with the simplest case of keeping the scalar field potential zero, i.e., $m_\varphi = 0 = \lambda$, and vary only the conformal factor via $\beta$; as explained before, this will lead to the largest deviations from pure general relativity ($m_\varphi > 0$ or $\lambda > 0$ will suppress the scalar field, see also Fig.~\ref{fig:M_R}). As the $f$-mode is an acoustic mode and hence its frequency is closely linked to the size of the star, we can---at least qualitatively---predict the change in frequency from the mass-radius diagram in Fig.~\ref{fig:M_R}; for simplicity, we focus on the most extreme case shown in red with upward triangles (and the precise results are shown in the left panel in Fig.~\ref{fig:f_MR3}): starting from the low density range of the mass-radius curve, the scalarization first pushes the equilibrium configurations toward smaller radii once we cross the first bifurcation point, hence the $f$-mode will increase in frequency when compared to the non-scalarized models. As we continue along the mass-radius curve toward compacter models (despite the large deviation from the GR curve, the average density keeps monotonically increasing), the scalar field flips its effect and increases the neutron star's radius (and decreases its average density), coming along with a decrease in frequency; this continues up until the upper bifurcation point. Note that this description is of purely qualitative nature and the approximately linear dependence of the $f$-mode frequency on the (square root of) star's average density that has been observed in GR no longer holds in STT. The effect of the scalar field is nearly negligible for $\beta = -4.5$ but becomes increasingly pronounced for smaller $\beta$ (we considered models down to $\beta = -6$). 

Let us now investigate the impact of a non-vanishing scalar potential. For this, we keep $\beta = -6$ fixed and vary the other two parameters $m_\varphi$ and $\lambda$; in order to separate the effects of those two, we will vary only one at a time.

First, we will focus on the middle panel of Fig.~\ref{fig:f_MR3} where we increase $\lambda$. There, we show the $f$-mode frequency as a function of the average density of the star and the red line with triangles displays the largest deviation from general relativity. As we increase $\lambda$, the $f$-mode frequencies move closer to the frequencies as obtained in pure general relativity; the difference to pure GR is only marginal for $\lambda = 1$ and for $\lambda = 10$ it is indistinguishable in the diagram. This is due to the fact that an increased value of $\lambda$ suppresses the scalar field and hence the scalarized equilibrium configurations are closer to the purely general relativistic case than if $\lambda$ was zero.

Next, we consider the impact of a massive scalar field by increasing $m_\varphi$ (and setting $\lambda = 0$ again), cf. the right panel of Fig.~\ref{fig:f_MR3}. We observe a similar effect to that of a non-zero $\lambda$: with increasing scalar field mass, the frequencies move closer to the general relativistic values. While the impact of the scalar field's mass is rather small for $m_\varphi = 0.005$, the $f$-mode frequency is nearly indistinguishable from the GR values for $m_\varphi = 0.05$. However, the massive scalar field also impacts the bifurcation points between which scalarized equilibrium configurations different from the GR solution exist. The more massive the scalar field is, the smaller the parameter window for scalarized neutron stars becomes.

It is natural to expect that an additional class of modes will emerge in the spectrum which is associated with the scalar field and this class was indeed identified in studies concerning radial oscillations \cite{Mendes:2018qwo,Blazquez-Salcedo:2020ibb,Doneva:2020csi}. It is a challenging task to extract these modes from a time evolution, though, since they have a rather short damping time and even if they are present in the signal, they are damped already after one or two oscillation cycles to such low amplitudes that they are essentially swallowed by the fluid oscillations. Particular initial data are required that excite the scalar modes to sufficiently large amplitude while keeping the fluid oscillations low so that they can be distinguished for a longer period of time; nonetheless, the coupling to the fluid will inevitably set it in motion, leading to the same intricacy as just described. Even though we had some indications that such modes are present in our simulations, we were not able to perform a clear and unambiguous identification of those modes at this time. Since the main focus of our paper is the calculation of the fluid polar modes, we leave their further investigation for future studies.

\subsection{Accuracy of the Cowling approximation}
\label{sec:cowling}

At the end, we will discuss in detail the accuracy of the Cowling approximation for the fundamental $\ell=2$ $f$-mode oscillations of the neutron stars in order to know how accurate qualitatively and quantitatively the previous results on the subject are \cite{Sotani:2004rq,Staykov:2015cfa} and to be able to give a rough prediction whether the Cowling approximation is justified for other alternative theories theories of gravity, where the corresponding perturbation equations can be much more complicated.

The Cowling approximation has been extensively used for studies in GR (see, e.g., \cite{1990ApJ...348..198L,Gaertig:2008uz,Doneva:2013zqa,Sotani:2020mwc} and references therein) and it can overestimate the $f$-mode frequencies up to $30\%$. The qualitative behavior of the modes does not change, though, only the frequencies are shifted (increased) with respect to the true results, even in the case of rapid rotation \cite{Doneva:2013zqa,Kruger:2019zuz,Kruger:2020ykw}.

There is no unique definition of \emph{the} Cowling approximation in alternative theories of gravity. The most straightforward (and simplest from computational point of view) is to assume that both the spacetime and the scalar field perturbations are zero. We will call this ``full Cowling'' and this is the most studied case because it leads not only to simpler equations, but also the computational domain is limited to the interior of the star. This approach was undertaken for example in \cite{Sotani:2004rq,Staykov:2015cfa,Silva:2014ora}. One can go one step further and allow for the evolution of the scalar field while keeping the metric fixed. This approach was performed in \cite{Sotani:2014tua} and we will name it ``classic Cowling''. The last option is to keep the scalar field fixed while evolving the metric, which we will call ``scalar Cowling''.

The comparison between the PSD in the case without approximation and for ``scalar Cowling'' is presented in the right panel of Fig. \ref{fig:timepsd1} for two representative scalarized neutron star models while the numerical values of the extracted frequencies are shown in Tab. \ref{tab:freq}. Quite similar to the Cowling approximation in GR, freezing the scalar field to its equilibrium value also pushes the fluid frequencies to higher values. The impact on the frequencies is in general weak, though, with the strongest deviation reached for the $f$-mode. A comparison between the PSD in the case without approximation and for ``classic Cowling'' is presented in the right panel of Fig. \ref{fig:timepsd2}.  As before, and as well known from the Cowling approximation in purely general relativistic studies, the frequencies of the acoustic modes are shifted to larger values. In contrast to the ``scalar Cowling'' case before, where we kept the scalar field fixed but evolved the spacetime, the impact of this ``classic Cowling'' approximation is stronger.

\begin{figure}
	\includegraphics[width=1\textwidth]{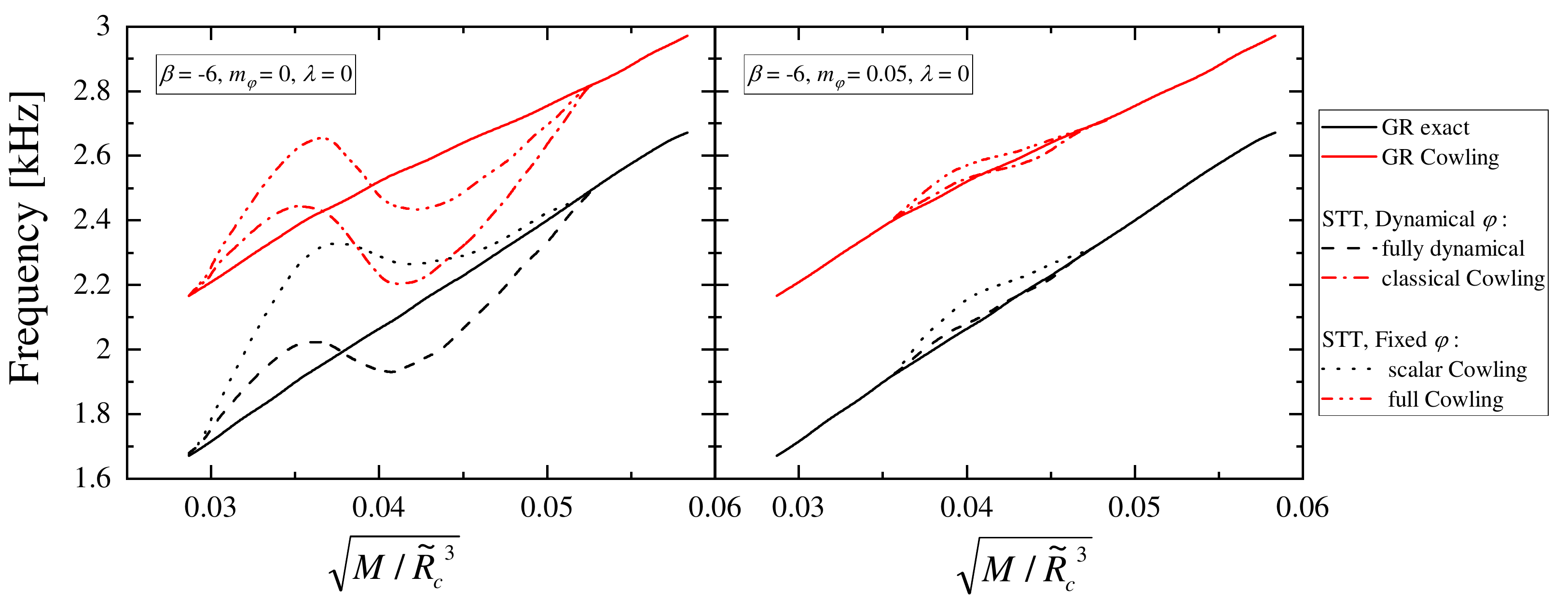}
	\caption{Comparison of $f$-mode frequencies for several series of models with varying $\epsilon_c$, where the space-time, scalar field or both or none of those are held fixed. The three blue curves have a dynamic space-time while the red curves represent a frozen space-time. The two solid lines are the purely general relativistic case ($\beta = 0$), whereas the dashed (scalar field dynamic) and dash-dotted (scalar field fixed) curves are obtained with $\beta = -6$ (and $m_\varphi = \lambda = 0)$. }
	\label{fig:cowling_comp}
\end{figure}

To visualize the effect of the different Cowling variants defined above, we show the $f$-mode frequency for two representative combinations of scalar field parameters in Fig.~\ref{fig:cowling_comp}; the pure GR values are shown with solid lines for comparison.  The left panel represents the case with $\beta=-6$ and vanishing scalar field potential, while in the right one we have relatively high scalar field mass reducing significantly the deviations from GR and shrinking the domain of existence of scalarized solutions ($\beta=-6$, $m_\varphi=0.05$ and $\lambda=0$). Looking at the right panel first, it is obvious that the Cowling variant which induces the smallest absolute deviation from the fully dynamic situation (shown with dashed lines) is the ``scalar Cowling'' approximation (dotted lines), as expected. Both the ``classic Cowling'' and the ``full Cowling'' approximations (both shown in red) in which the spacetime is held fixed yield a qualitatively very similar picture compared to the cases with a dynamic spacetime (in black). To a good approximation the ``fixed spacetime'' (red) sequences are merely moved to higher frequencies with respect to the ``dynamical spacetime'' branches. The shift appears to be quite similar for the whole range of average compactness and it is of the same order as the shift between the Cowling (solid red) and full (solid black) GR sequences.

Next, in the case of vanishing scalar field potential (left panel of Fig. \ref{fig:cowling_comp}), the deviations from GR induced by the presence of a dynamic scalar field are qualitatively similar in both cases of a dynamic and static spacetime. There, the parameter window in which scalarization occurs is split into two parts: one of lower average density in which the $f$-mode frequency is increased with respect to GR, while it is decreased in the one of higher average density. The picture changes slightly when we compare the GR case to the scalar Cowling approximation; here, the $f$-mode frequency of the scalarized neutron stars is increased compared to GR for all average densities; nonetheless, the feature of a local maximum and a local minimum in the frequency still occurs.  Again, the impact of the spacetime being frozen can be very roughly approximated by a constant frequency shift with respect to the fully dynamical results and also in this case the shift is quite similar to the difference between the Cowling (solid red) and full (solid black) GR sequences.

Thus, we can provide a simple recipe to estimate the frequency of the $f$-mode in the fully dynamic case even without evolving the scalar field or the spacetime (which is coupled to the scalar field) in time: first, calculate the spectrum in the ``full Cowling'' approximation, which should in general a comparatively feasible task; second, subtract from those values the difference between purely general relativistic value and its Cowling counterpart---it is well-known how to calculate this quantity. This simple calculation should give a useful (even quantitatively!) estimate for the $f$-mode frequency in the fully dynamic case.

It is highly probable, that similar conclusions will hold also for other scalar-tensor types of gravitational theories. Since the field equations can be extremely complicated for such theories, having the opportunity to obtain qualitatively good results and even some quantitative estimation by adopting some form of a Cowling approximation, is of great value.

\section{Conclusions}
In the present paper we study the oscillation modes of spontaneously scalarized neutron stars. We consider scalar-tensor theories with and without scalar field potential, including both a massive and a self-interacting term. For this purpose the perturbation equations governing the neutron star oscillations are derived without approximations contrary to the previous studies on this subject. A numerical code is developed for the solution of these equations in the time domain. Even though this approach is typically somewhat inferior in accuracy compared to the solution of the equations in the frequency domain, it still leads to quantitatively accurate results, where the accuracy is of the order of 1-2\,\%. More importantly, it is much more tractable and straightforward to handle.

We have calculated the oscillation frequencies of a large number of models spanning a wide range of the free parameters; these are the STT coupling constant $\beta$, the mass of the scalar field $m_\varphi$ and the self-interaction parameter $\lambda$. For all of these models we have calculated the fundamental $\ell=2$ $f$-mode and its first two overtones. As expected from the behavior of the background equilibrium solutions,  the decrease of $\beta$ leads to an increase of the deviations in the oscillations frequencies compared to pure GR. Nonzero $m_\varphi$ or $\lambda$ suppress the scalar field and for large values they lead to an oscillation spectrum that is practically indistinguishable from GR. 

It is well known that in GR the frequencies of the $f$-modes scale linearly with the (square root of) average density of the star. We demonstrate that this is not fulfilled in scalar-tensor theories of gravity where the deviation from a linear dependence can be significant. It might be possible to introduce a new type of universal asteroseismology relations with a proper normalization of the parameters which holds also for scalarized neutron stars; such a study is currently in progress.

The previous studies in the field were done mainly with the use of some kind of Cowling approximation, assuming that the spacetime metric and/or the scalar field are fixed. The applicability and accuracy of this approximation was evaluated only in pure general relativity where it was shown that it overestimates the $f$-mode frequency by up to 30\%; despite this deviation, it preserves the qualitative behavior of the $f$-mode frequency well. In scalar-tensor theories, different variants of the Cowling approximation can be defined and we have demonstrated that all of them produce to some good extent the same qualitative behaviour of the $f$-mode frequencies. We find that even some quantitative estimates for the true frequencies (i.e., in the fully dynamic case) can be made based on the results in the classic Cowling approximation: the difference between a dynamic and a frozen spacetime is to good approximation independent of whether we consider a purely general relativistic neutron star or a scalarized neutron star (with or without scalar potential). We hypothesize, that such type of approximations can be very useful to track the behavior of the neutron star oscillation spectrum and the related gravitational wave emission in other alternative theories of gravity, where the solution of the full perturbation equations is considerably more involved.

\begin{acknowledgments}
We would like to thank Stoytcho Yazadjiev for the continuous support and guidance as well as many fruitful discussions in the preparation of this work. DD acknowledges financial support via an Emmy Noether Research Group funded by the German Research Foundation (DFG) under Grant no. DO 1771/1-1  and the partial support by the National Science
Fund of Bulgaria under Contract No. K-06-N-38/12. DD is indebted to the Baden-W\"urttemberg Stiftung for the financial support of this research project by the Eliteprogramme for Postdocs. SY would like to thank the University of T\"ubingen for the financial support. The partial support by the Bulgarian NSF Grant KP-06-H28/7 and the Networking support
12 by the COST Actions CA16104 and CA16214 are also gratefully acknowledged. CK acknowledges financial support through DFG research Grant No. 413873357.
\end{acknowledgments}

\begin{widetext}

\appendix

\section{Definitions of the Perturbation Variables}
\label{app:def_pert}

In the derivation of the evolution equations, we follow closely the approach taken in \cite{Kruger:2020ykw},  to which we refer the reader for further details. However, in the present study, we limit ourselves to non-rotating stars  which simplifies the expressions significantly.

The spherically symmetric perturbation problem naturally decomposes into two independent subsets: polar and axial perturbations. As we are interested in polar perturbations only, three of the ten metric perturbations vanish. We define the other seven to be
\begin{align}
    \phi_{tt}
        & = -2e^{2\nu}\mcH, \label{eq:metricpert_tt} \\
    \phi_{tr}
        & = \mcL, \label{eq:metricpert_tr} \\
    \phi_{t\theta}
        & = r \mcM, \label{eq:metricpert_tth} \\
    \phi_{rr}
        & = 2 e^{2\mu} \mcK, \label{eq:metricpert_rr} \\
    \phi_{r\theta}
        & = e^{2\mu} r \mcQ, \label{eq:metricpert_rth} \\
    \phi_{\theta\theta}
        & = 2 e^{2\mu} r^2 \mcP, \label{eq:metricpert_thth} \\
    \phi_{\hat{\varphi}\hat{\varphi}}
        & = 2 e^{2\psi} \rrsst \mcW, \label{eq:metricpert_phph}
\end{align}
i.e., we have introduced the 7 perturbations $\mcH, \mcK, \mcL, \mcM, \mcP, \mcQ, \mcW$. Note that we use the hat in order to distinguish the azimuthal spherical coordinate $\hat{\varphi}$ from the scalar field $\varphi$. For the fluid quantities, we use the components of the ``Cowling part'' $\delta T^{\mu\nu}_C$ of the perturbed energy-momentum tensor $\delta T^{\mu\nu}$, both of which are related via
\begin{align}
    \delta T^{\mu\nu}
        & =: \delta T^{\mu\nu}_C - p h^{\mu\nu}.
\end{align}
The fluid perturbations then are
\begin{align}
    Q_1 & := \delta T^{tt}_{\rm C},        \label{eq:def_Q1} \\
    Q_3 & := \delta T^{tr}_{\rm C},        \label{eq:def_Q3} \\
    Q_4 & := \delta T^{t\theta}_{\rm C} r, \label{eq:def_Q4} \\
    Q_5 & := \delta T^{\hat{\varphi}\hat{\varphi}}_{\rm C},  \label{eq:def_Q5} \\
    Q_6 & := \delta T^{rr}_{\rm C} r^2 \sin^2 \theta.        \label{eq:def_Q6}
\end{align}
By means of these definitions, it can easily be seen that $Q_5 = Q_6$ (in the non-rotating case). Furthermore, we will see later that $Q_6$ is not independent, either.

The perturbation of the scalar field will simply be $\delta\varphi$. Summarizing, we will need to evolve eleven perturbation quantities in time: seven for the spacetime, three for the fluid and one for the scalar field.

We exploit the spherical symmetry of our problem to remove the angular dependence of our perturbation variables and express their angular dependence in suitable combinations of the spherical harmonics $Y_{lm}(\theta, \hat{\varphi})$. We use the subscript ``${}_0$'' to denote the newly defined perturbation variables that will depend on $t$ and $r$ only. We have the following decompositions into spherical harmonics
\begin{align}
    X
        & = X_0 r^{\ell-2} Y_{lm}
    & & \text{for}\quad X \in \{ \mcL, Q_1, Q_3, Q_6, \delta\varphi \}, \\
\intertext{and}
    X
        & = X_0 r^{\ell-2} \partial_\theta Y_{lm}
    & & \text{for}\quad X \in \{ \mcM, \mcQ, Q_4 \}.
\end{align}
The remaining four perturbation variables $\mcH_0$, $\mcK_0$, $\mcP_0$, and $\mcW_0$ are defined via the linear combinations
\begin{align}
    \mcH
        & = \frac{1}{4}
            \left[
                \mcH_0 - \mcK_0 - 2 \mcP_0 + \ell(\ell+1) \mcW_0
            \right] r^{\ell-2} Y_{lm},
    \\
    \mcK
        & = \frac{1}{4}
            \left[
                - \mcH_0 + \mcK_0 - 2 \mcP_0 + \ell(\ell+1) \mcW_0
            \right] r^{\ell-2} Y_{lm},
    \\
    \mcP
        & = - \frac{1}{4}
            \left[ \left(\mcH_0 + \mcK_0 \right) Y_{lm}
            + \left[
                2 \cot\theta \partial_\theta Y_{lm}
                + \ell(\ell+1) Y_{lm}
              \right] \mcW_0
            \right] r^{\ell-2},
    \\
    \mcW
        & = - \frac{1}{4}
            \left[ \left(\mcH_0 + \mcK_0 \right) Y_{lm}
            - \left[
                2 \cot\theta \partial_\theta Y_{lm}
                + \ell(\ell+1) Y_{lm}
              \right] \mcW_0
            \right] r^{\ell-2}.
\end{align}

The factor of $r^{\ell-2}$ in these definitions corresponds to the natural behaviour of the perturbation variables at the origin $r=0$ (as discovered via a Taylor expansion) and ensures regularity of the solution there.

\section{The Evolution Equations}
\label{app:evol_eq}

In writing down the evolution equations, we use the comma notation to abbreviate partial derivatives of the background quantities, for example, $\psi_{,r} := \partial_r \psi$, and we use the shortcut $\kappa := 4 \pi e^{2\psi} (\epsilon + p)$.

The Hilbert condition (cf. Eq. \eqref{eq:hilbert_gauge}) yields three equations:
\begin{align}
    e^{2\psi} \pder{t} \mcH_0
        & = - 2\left(\nu_{,r} + \psi_{,r} + \frac{\ell}{r} \right) \mcL_0
            + \frac{2 \ell (\ell+1)}{r} \mcM_0
        \nonumber \\
        & \qquad
            + e^{2\psi} \pder{t} \mcK_0
            - 2 \pder{r} \mcL_0
            + 2 e^{2\psi} \pder{t} \mcP_0
            - e^{2\psi} \ell ( \ell + 1 ) \pder{t} \mcW_0, \\
    e^{-2\nu} \pder{t} \mcL_0
        & = - \left( \nu_{,r} + \frac{\ell - 2}{2r} \right) \mcH_0
            + \left( \nu_{,r} + 2 \psi_{,r} + \frac{\ell + 2}{2r} \right) \mcK_0
            - \left( 2\psi_{,r} + \frac{\ell}{r} \right) \mcP_0
        \nonumber \\
        & \qquad
            - \frac{\ell(\ell+1)}{r} \mcQ_0
            + \frac{\ell(\ell+1)}{2r} \left( 2\psi_{,r} + \ell \right) \mcW_0
            - \frac{1}{2} \pder{r} \mcH_0
            + \frac{1}{2} \pder{r} \mcK_0
        \nonumber \\
        & \qquad
            - \pder{r} \mcP_0
            + \frac{\ell(\ell+1)}{2r} \pder \mcW_0, \\
    e^{-2\nu} \pder{t} \mcM_0
        & = - \frac{1}{2r} \mcH_0
            - \frac{1}{2r} \mcK_0
            + \left( \nu_{,r} + 3\psi_{,r} + \frac{\ell+1}{r} \right) \mcQ_0
        \nonumber \\
        & \qquad
            - \frac{(\ell+2)(\ell-1)}{2r} \mcW_0
            + \pder{r} \mcQ_0.
\end{align}
Since the Hilbert gauge is a purely general relativistic construct, the Hilbert conditions are unaltered by the presence of a scalar field. These three Hilbert conditions may be used to simplify the evolution equations or to monitor the violation of the Hilbert gauge throughout the time evolution.

The perturbed field equations (cf. Eq.~\eqref{eq:fieldeq}) provide us with seven wave equations for the spacetime perturbations:
\begingroup
\allowdisplaybreaks
\begin{align}
    e^{2\psi-2\nu} \ppder{t} \mcL_0
        & = - 16 \pi e^{4\psi+2\nu} Q_{30}
            - \left[ 2 \varphi_{,r}^2 + 3 \nu_{,r}^2 + 2 \psi_{,r} \nu_{,r} + \psi_{,r}^2
                + \frac{1}{r} (\nu_{,r} + \psi_{,r}) (\ell+1)
                + \frac{4\ell}{r^2}
              \right] \mcL_0
        \nonumber \\
        & \qquad
            - \left( \nu_{,r} + \psi_{,r} - \frac{2\ell-2}{r} \right) \pder{r} \mcL_0
            + \ppder{r} \mcL_0
            + \frac{2\ell(\ell+1)}{r} \left( \psi_{,r} + \frac{1}{r} \right) \mcM_0
        \nonumber \\
        & \qquad
            + 2 e^{2\psi} \nu_{,r} \pder{t} \mcK_0
            - 2 e^{2\psi} \nu_{,r} \pder{t} \mcH_0
            + 4 e^{2\psi} \varphi_{,r} \pder{t} \delta\varphi_0 \\
    e^{2\psi-2\nu} \ppder{t} \mcM_0
        & = - 16 \pi e^{4\psi+2\nu} Q_{40}
            + \frac{2}{r} \left( \psi_{,r} + \frac{1}{r} \right) \mcL_0
        \nonumber \\
        & \qquad
            + \left[ 2 \varphi_{,r}^2 - 2 \kappa - 2 \psi_{,r}^2
                + \frac{1}{r} \left( \nu_{,r}(\ell-1) - \psi_{,r}(\ell+3)  \right)
                - \frac{4\ell-2}{r^2}
              \right] \mcM_0
        \nonumber \\
        & \qquad
            - \left( \nu_{,r} + \psi_{,r} - \frac{2\ell-2}{r} \right) \pder{r} \mcM_0
            + 2 e^{2\psi} \nu_{,r} \pder{t} \mcQ_0
            + \ppder{r} \mcM_0, \\
    e^{2\psi-2\nu} \ppder{t} \mcQ_0
        & = \frac{2}{r} \left( \psi_{,r} + \frac{1}{r} \right) \mcK_0
            - \frac{2}{r} \left( \psi_{,r} + \frac{1}{r} \right) \mcP_0
        \nonumber \\
        & \qquad
            - \left[ 4 \varphi_{,r}^2 + 2 \kappa + \nu_{,r}^2 - 2 \psi_{,r} \nu_{,r} + 5 \psi_{,r}^2
                - \frac{1}{r} \left( \nu_{,r} \ell + \psi_{,r}(\ell-12)  \right)
                + \frac{4\ell+2}{r^2}
              \right] \mcQ_0
        \nonumber \\
        & \qquad
            + \left( \nu_{,r} + \psi_{,r} + \frac{2\ell-2}{r} \right) \pder{r} \mcQ_0
            + \ppder{r} \mcQ_0
            + \frac{2 (\ell^2 + \ell -1)}{r} \left( \psi_{,r} + \frac{1}{r} \right) \mcW_0
        \nonumber \\
        & \qquad
            + 2 e^{-2\nu} \nu_{,r} \pder{t} \mcM_0
            + \frac{4}{r} \varphi_{,r} \delta\varphi_0, \\
    e^{2\psi-2\nu} \ppder{t} \mcH_0
        & = - 8 \pi e^{2\psi+2\nu} Q_{10} - 24 \pi e^{4\psi} Q_{60}
            - \left[ 2 \kappa + 2 \nu_{,r}^2
                - \frac{1}{r} (\nu_{,r} + \psi_{,r}) (\ell-2)
                + \frac{4\ell-2}{r^2}
              \right] \mcH_0
        \nonumber \\
        & \qquad
            + \left( \nu_{,r} + \psi_{,r} + \frac{2\ell-2}{r} \right) \pder{r} \mcH_0
            + \ppder{r} \mcH_0
            + 2\left[ \varphi_{,r}^2 - \kappa + \nu_{,r}^2 - \psi_{,r}^2 - \frac{2}{r} \psi_{,r}
              \right] \mcK_0
        \nonumber \\
        & \qquad
            + 4 \nu_{,r} \left( \psi_{,r} + \frac{1}{r} \right) \mcP_0
            + 2 \nu_{,r} \ell (\ell+1) \left( \psi_{,r} + \frac{1}{r} \right) \mcW_0
            - 4 e^{-2\nu} \nu_{,r} \pder{t} \mcL_0
        \nonumber \\
        & \qquad
            + 4 e^{2\psi} \frac{\partial V(\varphi)}{\partial\varphi} \delta \varphi_0, \\
    e^{2\psi-2\nu} \ppder{t} \mcK_0
        & = 8 \pi e^{2\psi+2\nu} Q_{10} - 8 \pi e^{4\psi} Q_{60}
            + 2 \left[ \varphi_{,r}^2 + \nu_{,r}^2 + \psi_{,r}^2
                - \frac{2}{r} \psi_{,r}
              \right] \mcH_0
        \nonumber \\
        & \qquad
            - \left[ 4 \varphi_{,r}^2 + 2 \nu_{,r}^2 + 4 \psi_{,r}^2
                - \frac{1}{r} \left( \nu_{,r} (\ell-2) + \psi_{,r}(\ell-10)  \right)
                + \frac{4\ell+2}{r^2}
              \right] \mcK_0
        \nonumber \\
        & \qquad
            + \left( \nu_{,r} + \psi_{,r} + \frac{2\ell-2}{r} \right) \pder{r} \mcK_0
            + \ppder{r} \mcK_0
            + \frac{4 \ell (\ell+1)}{r} \left( \psi_{,r} + \frac{1}{r} \right) \mcQ_0
        \nonumber \\
        & \qquad
            + 4 \left[ \varphi_{,r}^2 + \psi_{,r}^2 - \psi_{,r} \nu_{,r} + \kappa
                - \frac{1}{r} \left( \nu_{,r} - 2 \psi_{,r} \right)
                + \frac{1}{r^2}
              \right] \mcP_0
        \nonumber \\
        & \qquad
            - \frac{2\ell(\ell+1)}{r} \left[ \varphi_{,r}^2 + \psi_{,r}^2 - \psi_{,r} \nu_{,r} + \kappa
                - \frac{1}{r} \left( \nu_{,r} - 2 \psi_{,r} \right)
                + \frac{1}{r^2}
              \right] \mcW_0
        \nonumber \\
        & \qquad
             + 4 e^{-2\nu} \nu_{,r} \pder{t} \mcL_0
             + 4 \left[ e^{2\psi} \frac{\partial V(\varphi)}{\partial\varphi}
                + \frac{2\ell -4}{r} \varphi_{,r}
                \right] \delta\varphi_0
             + 8 \varphi_{,r} \pder{r} \delta\varphi_0, \\
    e^{2\psi-2\nu} \ppder{t} \mcP_0
        & = 8 \pi e^{2\psi+2\nu} Q_{10} - 8 \pi e^{4\psi} Q_{60}
            + 2 \left[ \kappa - \psi_{,r} \nu_{,r} - \frac{1}{r} \nu_{,r}
              \right] \mcH_0
        \nonumber \\
        & \qquad
            + 2 \left[ \varphi_{,r}^2 + \psi_{,r}^2 - \psi_{,r} \nu_{,r} + \kappa
                - \frac{1}{r} \left( \nu_{,r} - 2 \psi_{,r} \right)
                + \frac{1}{r^2}
              \right] \mcK_0
        \nonumber \\
        & \qquad
            - \left[ 4 \psi_{,r}^2
                - \frac{1}{r} \left( \nu_{,r}(\ell-2) + \psi_{,r}(\ell-10) \right)
                + \frac{4\ell}{r^2}
              \right] \mcP_0
        \nonumber \\
        & \qquad
            + \left( \nu_{,r} + \psi_{,r} + \frac{2\ell-2}{r} \right) \pder{r} \mcP_0
            + \ppder{r} \mcP_0
            + 2 \ell (\ell+1) \left( \psi_{,r} + \frac{1}{r} \right)^2 \mcW_0
        \nonumber \\
        & \qquad
            + 4 e^{2\psi} \frac{\partial V(\varphi)}{\partial\varphi} \delta\varphi_0, \\
    e^{2\psi-2\nu} \ppder{t} \mcW_0
        & = \frac{4}{r} \left( \psi_{,r} + \frac{1}{r} \right) \mcQ_0
            + \frac{1}{r} \left[
                (\nu_{,r} + \psi_{,r}) (\ell-2)
                - \frac{4\ell-4}{r}
            \right] \mcW_0
        \nonumber \\
        & \qquad
            + \left( \nu_{,r} + \psi_{,r} + \frac{2\ell-2}{r} \right) \pder{r} \mcW_0
            + \ppder{r} \mcW_0.
\end{align}
\endgroup
As can be seen, the presence of a scalar field leads to only small modifications of the perturbation equations when compared to the purely general relativistic case: a few coefficients gain an additional $\varphi_{,r}^2$ term and most equations recruit a source term due to the scalar field perturbation $\delta\varphi_0$.

The conservation of energy-momentum, Eq.~\eqref{eq:convlaw}, results in three evolution equations for the hydrodynamics:
\begin{align}
    e^{2\nu} \pder{t} Q_{10}
        & = - e^{2\nu} \left( 3\nu_{,r} + 3\psi_{,r} + \frac{\ell}{r} \right) Q_{30}
            - e^{2\nu} \pder{r} Q_{30}
            + e^{2\nu} \frac{\ell(\ell+1)}{r} Q_{40}
        \nonumber \\
        & \qquad
            - \frac{1}{2} (\epsilon + p) \left( 2 \pder{t} \mcH_0 + \pder{t} \mcK_0 + 2 \pder{t} \mcP_0 - \ell(\ell+1) \mcW_0 \right)
        \nonumber \\
        & \qquad
            + \alpha(\varphi) (\epsilon - 3p) \pder{t} \delta\varphi_0, \\
    e^{2\psi} \pder{t} Q_{30}
        & = - e^{2\nu} \left[ \nu_{,r} + \varphi_{,r} \alpha(\varphi) \right] Q_{10}
            + e^{2\psi} \left(
                3 \varphi_{,r} \alpha(\varphi) - \nu_{,r} - \psi_{,r} - \frac{\ell-2}{r}
                \right) Q_{60}
        \nonumber \\
        & \qquad
            - e^{2\psi} \pder{r} Q_{60}
            - \frac{1}{2} (\epsilon + p)
                \left[
                    2 \varphi_{,r} \alpha(\varphi) + 2 \nu_{,r} + \frac{\ell-2}{r}
                \right] \mcH_0
            - \frac{1}{2} (\epsilon + p) \pder{r} \mcH_0
        \nonumber \\
        & \qquad
            - e^{-2\nu} (\epsilon + p) \pder{t} \mcL_0
            + (3p - \epsilon) \left[
                \frac{\ell-2}{r} \alpha(\varphi)
                + \varphi_{,r} \frac{\partial \alpha(\varphi)}{\partial\varphi}
                \right] \delta\varphi_0
        \nonumber \\
        & \qquad
            + \alpha(\varphi) (3p - \epsilon) \pder{r} \delta\varphi_0, \\
    e^{2\psi} \pder{t} Q_{40}
        & = - \frac{e^{2\psi}}{r} Q_{60}
            - \frac{1}{2r} (\epsilon + p) \mcH_0
            - e^{-2\nu} (\epsilon + p) \pder{t} \mcM_0
            + \frac{1}{r} \alpha(\varphi) (3p - \epsilon) \delta\varphi_0.
\end{align}
The presence of a dynamic scalar field is reflected in these three equations in a very similar way as it is in the wave equations for the spacetime perturbations.

Finally, the scalar field equation \eqref{eq:fieldeqscalar} yields an evolution equation for the perturbation of the scalar field:
\begin{align}
    e^{2\psi-2\nu} \ppder{t} \delta\varphi_0
        & =
            - 4 \pi e^{2\psi+2\nu} \alpha(\varphi) Q_{10}
            + 12 \pi e^{4\psi} \alpha(\varphi) Q_{60}
            - \left[
                4 \pi e^{2\psi} (\epsilon + p) \alpha(\varphi) + \nu_r \varphi_{,r}
            \right] \mcH_0
        \nonumber \\
        & \quad
            + \left[
                4 \pi e^{2\psi} \alpha(\varphi) (3p - \epsilon)
                - e^{2\psi} \frac{\partial V(\varphi)}{\partial\varphi}
                + (\nu_{,r} + 2 \psi_{,r}) \varphi_{,r}
                + \frac{2}{r} \varphi_{,r}
            \right] \mcK_0
        \nonumber \\
        & \quad
            - 2 \varphi_{,r} \left( \psi_{,r} + \frac{1}{r} \right) \mcP_0
            + \ell(\ell+1) \varphi_{,r} \left( \psi_{,r} + \frac{1}{r} \right) \mcW_0
        \nonumber \\
        & \quad
            + \left[
                4 \pi e^{2\psi} (3p - \epsilon) \frac{\partial \alpha(\varphi)}{\partial\varphi}
                - e^{2\psi} \frac{\partial^2 V(\varphi)}{\partial\varphi^2}
                + \frac{1}{r}
                    (\nu_{,r} + \psi_{,r}) (\ell-2)
                - \frac{4\ell-2}{r^2}
            \right] \delta\varphi_0
        \nonumber \\
        & \quad
            + \left[
                \nu_{,r} + \psi_{,r} + \frac{2\ell-2}{r}
            \right] \pder{r} \delta\varphi_0
            + \ppder{r} \delta\varphi_0
\end{align}

All perturbation equations are written in the Einstein frame. For brevity, we have also kept the fluid quantities $\epsilon$ and $p$ (and their combination $\kappa$) in this frame, even though we need to keep in mind that their Jordan frame equivalent is the one that is physically relevant.

In addition to the hyperbolic differential equations, we are equipped with one more algebraic equation that allows us to compute $Q_{60}$ out of some other perturbation quantities: the definition of the speed of sound, $\delta \tilde{p} = \tilde{c}_s^2 \delta \tilde{\epsilon}$, can (in terms of our perturbation variables) be expressed as
\begin{align}
    Q_{60}
        & = e^{-2\psi} \tilde{c}_s^2 
            \left[ 
                e^{2\nu} Q_{10}
                + \left(\epsilon + p \right) \mcH_0
            \right]
            + 4 e^{-2\psi}
              \left(p - \tilde{c}_s^2 \epsilon \right) \alpha(\varphi) \delta\varphi_0.
\end{align}
The scalar field perturbation $\delta\varphi_0$ enters this relation since we have to translate the definition of the speed of sound from the Jordan frame (where it holds) to the Einstein frame (in which we perform our time evolution); this results in the relation $\delta p = \tilde{c}_s^2 \delta\epsilon + 4 (p - \tilde{c}_s^2 \epsilon) \alpha(\varphi) \delta\varphi_0$.

\end{widetext}

\bibliography{BIB/references}

\end{document}